\documentclass[aps,pra,twocolumn,groupaddress,showpacs,floatfix]{revtex4}
\usepackage{graphicx}

\begin{document}

\title{Phase diagram of harmonically confined one-dimensional fermions 
with attractive and repulsive interactions}

\author{V. L. Campo Jr.}
\author{K. Capelle}
\email{capelle@if.sc.usp.br}
\affiliation{Departamento de F\'{\i}sica e Inform\'atica,
Instituto de F\'{\i}sica de S\~ao Carlos,
Universidade de S\~ao Paulo,
Caixa Postal 369, 13560-970 S\~ao Carlos, SP, Brazil}
\date{\today}

\begin{abstract}
We construct the complete $U$-$\mu$ phase diagram for harmonically confined
ultracold fermionic atoms with repulsive and attractive interactions.
($\mu$ is the chemical potential and $U$ the interaction strength.)
Our approach is based on density-functional theory, and employs analytical
expressions for the kinetic and correlation energy functionals. 
For repulsive interactions our calculations confirm previous 
numerical studies, and complement these by closed expressions for 
all phase boundaries and characteristic lines of the phase diagram, and by
providing an explanation and solution for difficulties encountered in earlier 
density-functional work on the same system. For attractive interactions we
propose a new and accurate interpolation for the correlation energy, and
use it to extend the phase diagram to $U<0$.
\end{abstract}


\maketitle

\newcommand{\be}{\begin{equation}}
\newcommand{\ee}{\end{equation}}
\newcommand{\bea}{\begin{eqnarray}}
\newcommand{\eea}{\end{eqnarray}}
\newcommand{\bi}{\bibitem}

\renewcommand{\r}{({\bf r})}
\newcommand{\rp}{({\bf r'})}

\newcommand{\ua}{\uparrow}
\newcommand{\da}{\downarrow}
\newcommand{\la}{\langle}
\newcommand{\ra}{\rangle}
\newcommand{\dg}{\dagger}

The field of strongly correlated low-dimensional fermions has a long history 
and multifarious applications in condensed-matter physics.
More recent experimental progress in the field of confined ultracold atoms
\cite{atoms1,atoms2,atoms3,atoms4} has further widened the scope of this 
already very rich field by introducing, e.g., an unprecedented degree of 
control over the confining potential, and the possibility to continuously 
tune the particle-particle interaction from repulsive to attractive --- which 
is not easy for cold atoms, but unconceivable for electrons. After much 
progress on bosonic atoms and their condensation, more recently fermionic
atoms have come under intense study.

In the present paper we focus on one particular model for optically confined 
fermionic atoms, the one-dimensional Hubbard model. The parameters 
characterizing this model and spanning its phase diagram are the on-site 
interaction $U$ \cite{footnote0} and the filling factor $n=N/L$, where $N$ 
is the number of particles and $L$ the number of lattice sites. For systems 
in which $n$ is spatially varying, such as in the presence of a confining 
potential, it is more useful to specify the chemical 
potential $\mu$ instead, which is a constant throughout the system. Hence,
we are interested in the phase diagram of the one-dimensional Hubbard model
in $U$-$\mu$ space. Our aim is to map out this phase diagram, identify the
physics governing each possible phase, and to provide analytical expressions
for all phase boundaries.

The $U>0$, i.e., repulsive, part of this phase diagram has recently been 
studied also by Liu et al. \cite{drummond}, who used a half-numerical 
half-analytical local-density approximation (LDA).
Rigol et al. \cite{rigol} obtained essentially the same 
$U>0$ phase diagram from fully numerical Quantum Monte Carlo simulations.
An interesting feature that emerged in these works is the appearance of mixed
Mott-insulator Luttinger-liquid phases, characterized by spatially 
separated incompressible and compressible regions in the optical trap,
and plateaus in the density profiles coinciding with the incompressible regions.
Xianlong et al. \cite{polini} have recently studied the case of negative 
$U$, i.e., attractive interactions, where the physics is dominated by
Luther-Emery-type pairing correlations and the spontaneous formation of
atomic density waves. However, the Kohn-Sham-type density-functional
calculations of Ref. \cite{polini} do not allow one to map out the entire
phase diagram, because the self-consistency cycle does not converge in the
mixed metal-insulator phases at $U>0$. 

In the present paper we use an analytical local-density approximation both
for the kinetic energy (in the spirit of Thomas-Fermi theory) and for the
correlation energy (in the spirit of the Bethe-Ansatz LDA proposed in 
Ref.~\cite{balda}) to obtain the $U$-$\mu$ phase diagram. For repulsive 
interactions our results coincide with those obtained numerically
in Refs.~\cite{drummond,rigol}. However, as our approach is
entirely analytical, we obtain closed expressions for the various phase 
boundaries, which previously had to be extracted numerically. Moreover, our
present work also covers the case of attractive interactions, $U<0$, thus
mapping out the entire physically relevant phase diagram. 

To make our paper self-contained we start by briefly describing and 
comparing the various approximation schemes employed. Density-functional 
theory (DFT) \cite{kohnrmp,dftbook} shows that the total energy of a many-body
system in an arbitrary potential can be written
\be
E[n]= T_s[n] + E_H[n] + V[n] + E_{xc}[n],
\ee
where $T_s$ is the kinetic energy of noninteracting particles, $E_H$ is
the Hartree energy (i.e., the mean-field approximation to the interaction
energy), $V$ is the potential energy (arising from the spatial distribution of 
the nuclei in solid-state and molecular applications of DFT, and from the 
optical confining potential for atoms in optical traps), and $E_{xc}$ is the
exchange-correlation energy \cite{footnote1}. The local-density approximation
to any energy component $F$ is defined by $F^{LDA}=\int dr f(n)|_{n \to n\r}$
in the continuum, and $F^{LDA}= \sum_i f(n)|_{n \to n_i}$ on a lattice,
where $f$ denotes the per-volume or per-site value of $F$ in a spatially
uniform (homogeneous) system.

Although DFT is mostly applied to {\it ab initio} electronic-structure 
calculations in solid-state physics and quantum chemistry 
\cite{dftbook,kohnrmp}, its formal framework is sufficiently general to permit 
application also to wide classes of model Hamiltonians \cite{valtercondmat}, 
and, in particular, to the Hubbard model \cite{balda,gs,gsn}. A Bethe-Ansatz
based LDA functional for the 
correlation energy of the one-dimensional Hubbard model was proposed
in \cite{balda} and applied to Mott insulating phases \cite{mottepl} and 
superlattice structures \cite{superl}. These works employed a 
parametrization of the per-site ground-state energy of the spatially
uniform Hubbard model, $e(n,U)$, which was constructed \cite{balda} to recover 
exactly known results at $U=0$, $U\to \infty$ and $n=1$, and to be close
to data obtained from numerical solution of the Bethe-Ansatz (BA) 
\cite{liebwu} equations inbetween \cite{footnote2}. For $ 0 \leq n \leq 1$
\be
e(n,U) = 
-\frac{2\beta(U)}{\pi}\sin\left(\frac{\pi n}{\beta(U)}\right)
\label{lsoc1}
\ee
and for $ 1 \leq  n \leq 2$
\be
e(n,U)= 
(n-1)U -\frac{2\beta(U)}{\pi}\sin\left(\frac{\pi(2-n)}{\beta(U)}\right) 
\label{lsoc2}
\ee
Here $\beta(U)$ is obtained from
\be
-\frac{\beta}{\pi}\sin\left(\frac{\pi}{\beta}\right) = -2\int_0^\infty
\:\frac{J_0(x)J_1(x)}{x(1 + e^{Ux/2})}\:dx =: -2 I(U),
\ee
and $J_m$ is the $m$'th order Bessel function. The difference between the
$n<1$ and the $n>1$ branch is responsible for a derivative discontinuity of 
$E_c$ and the opening of a Mott gap at $n=1$ \cite{liebwu,mottepl}.

To perform analytical calculations for attractive interactions, our first 
task is to extend this construction to $U<0$. 
The construction of a parametrization of the per-site energy
is not unique, but a convenient and accurate choice is 
\be
e(n,U) = \frac{Un}{2} - 4I(|U|)\sin\left(\frac{\pi n}{2}\right).
\label{uneg}
\ee
This expression by construction satisfies the following exactly known
properties of $e(n,U)$ for negative $U$:
(i) $e(n=0,U) = 0$,
(ii) $e(n=1,U) = -4\,I(|U|) + U/2$,
(iii) $e(n,U=0) = -(4/\pi)\sin(\pi n/2)$,
(iv) $e(n,U \to -\infty) \to Un/2$,
(v) $\partial e(n,U)/\partial n|_{n=1} = U/2$, and
(vi) $e(2-n,U) + U(n-1) = e(n,U)$.
Note that unlike (\ref{lsoc1}) and (\ref{lsoc2}), the $U<0$ expression
(\ref{uneg}) is particle-hole symmetric [property (iv)], and defined on only
one branch.

The Kohn-Sham approach treats $T_s$ exactly, by means of single-particle 
orbitals, and approximates $E_{c}$, e.g., by the BA-LDA of Ref.~\cite{balda}.
\bea
E^{LDA}_{KS}[n,U] = T_s[n] + V[n] + E_H[n,U] + \sum_i e_c(n_i,U),
\label{kslda}
\eea
where for a spin-unpolarized system $E_H[n,U]=(U/4)\sum_i^L n_i^2$, and 
$V[n] =\sum_i n_i V_i$ is the potential energy in the confining potential 
$V_i$. $t_s(n)=e(n,U=0)$ is the per-site kinetic energy of noninteracting 
particles, and the difference $e(n,U)-e_H(n,U)-t_s(n) =: e_c(n,U)$ is 
by definition the correlation energy.

The total-energy LDA (TLDA), employed below, uses an LDA both 
for the kinetic and the correlation energy, so that
\bea
E^{TLDA}[n] = \sum_i t_s(n_i) + V[n] + E_H[n] + \sum_i e_c(n_i)
\\
= \sum_i e(n_i,U) + V[n],
\label{tlda}
\eea
where the second equality follows because the Hartree energy is a local 
functional for an on-site interaction. Note that the TLDA approach is 
different from the Thomas-Fermi approximation (TFA), which also employs an 
LDA for $T_s$, but takes $E_c\equiv 0$.

We now proceed to characterize the phase diagram. To this end we employ the
TLDA, which is more accurate than the TFA, and unlike KS procedures can yield
closed analytical results. The density profile is obtained from minimizing
the energy functional under the constraint of fixed total particle number.
Hence
\be
\frac{\delta(E-\mu N)}{\delta n_i} = 
\frac{\partial e(n_i,U)}{\partial n_i} + V_i - \mu = 0.
\ee
From the above parametrizations (\ref{lsoc1}), (\ref{lsoc2}) and (\ref{uneg})
of $e(n,U)$ we obtain the Euler equations
\bea
-2\cos\left(\frac{\pi n_i}{\beta}\right) + V_i- \mu = 0 \hspace*{0.4cm}
0 \leq n_i < 1 \label{euler1}\\
U + 2\cos\left(\frac{\pi (2-n_i)}{\beta}\right) + V_i- \mu = 0\ \hspace*{0.4cm}
1 < n_i \leq 2 \label{euler2}
\eea
for $U>0$, and 
\be
\frac{U}{2} - 2\pi I(|U|)\cos\left(\frac{\pi n}{2}\right) + V_i - \mu = 0
\label{euler3}
\ee
for $U<0$. Note that at $n_i=1$ the expression $e(n,U)$ for $U>0$ is
not differentiable and no Euler equation is obtained. The three Euler 
equations (\ref{euler1}) - (\ref{euler3}) have a complex solution space 
with a rich structure, which we now proceed to analyse.

\begin{figure}
\centering
\includegraphics[height=80mm,width=65mm,angle=-90]{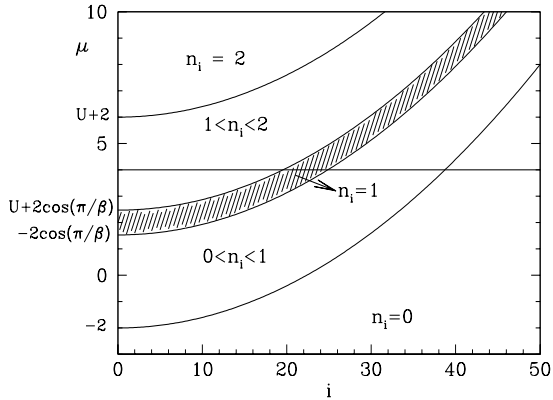}
\caption {\label{fig1} Classification of density shapes across the
$i>0$ half of a harmonic trap with $V_i = 0.004 i^2$, for $U=4>U^*$. 
The four dividing lines have the form $V_i + A$, where, from top to
bottom, $A=U+2$, $A=U+2 \cos(\pi/\beta)$, $A=- 2 \cos(\pi/\beta)$
and $A=-2$. The horizontal line at $\mu =4$ is the example discussed
in the main text.}
\end{figure}

Let us first consider $U>0$. In the low-density region far from the trap 
center the solution of Eq.~(\ref{euler1}) is
$n_i = \frac{\beta}{\pi}\arccos\left(\frac{V_i-\mu}{2}\right)$.
However, this solution only belongs to the interval $[0,1]$ if
\be
V_i - 2 \le \mu \le V_i- 2\cos\left(\pi/\beta\right).
\label{cond1}
\ee
Closer to the center of the trap $n_i>1$, and the relevant branch of the 
solution is (\ref{euler2}), leading to 
$n_i = 2-(\beta/\pi)\arccos((\mu - V_i-U)/2)$, provided that
\be
V_i + U + 2\cos\left(\pi/\beta\right) \le \mu \le V_i + U + 2.
\label{cond2}
\ee
Note that the conditions (\ref{cond1}) and (\ref{cond2}), which guarantee 
that the solution found does indeed pertain to the density interval it is 
obtained from, do not necessarily match continuously. In particular, if
\be
U+2\cos\left(\pi/\beta\right) > -2\cos\left(\pi/\beta\right)
\label{ustar}
\ee
there are values of $\mu$ for which 
neither of the two branches of the Euler equation has a solution. The actual 
density in this situation is then the one for which the Euler equations
are not defined, i.e., $n_i=1$. This possibility occurs only for $U>U^*$,
where $U^*=1.7349...$ is the lowest value of $U$ satisfying (\ref{ustar}).

\begin{figure}
\centering
\includegraphics[height=80mm,width=65mm,angle=-90]{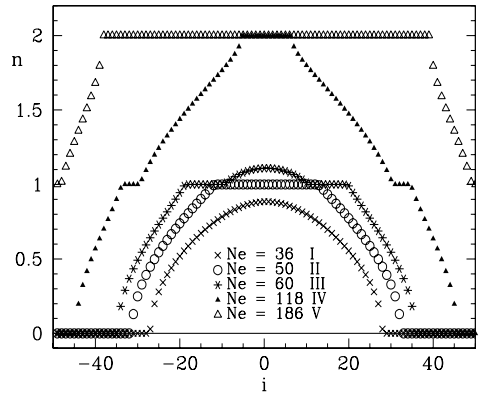}
\caption {\label{fig2} Typical density profiles for $U=4$ and a
confining potential $V_i=0.004 i^2$, for five different values of the
total fermion number, ilustrating the five basic types of density profiles
obtained for repulsive interactions.}
\end{figure}

\begin{figure}
\centering
\includegraphics[height=80mm,width=65mm,angle=-90]{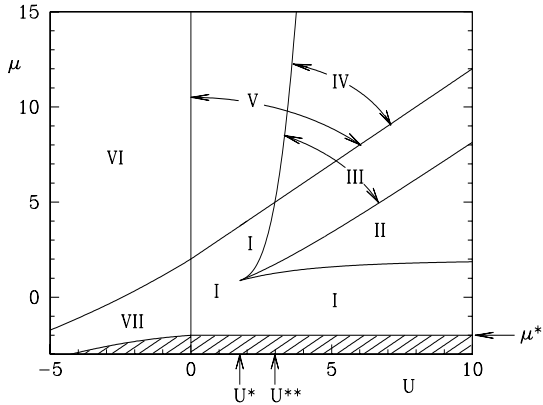}
\caption {\label{fig3} The $U$-$\mu$ phase diagram. See main text for details.}
\end{figure}

In Fig.~\ref{fig1} we plot the four limiting expressions defining the
boundaries of Eqs.~(\ref{cond1}) and (\ref{cond2}) for $U> U^*$. The chemical 
potential is constant throughout the trap, and thus corresponds to horizontal
lines crossing the various boundaries. As an illustration, consider a system
with $\mu =4$ (horizontal line in Fig.~\ref{fig1}). 
At the trap center ($i=0$) one first finds a high-density
solution with $1 < n_i < 2$. Moving outward, one crosses the left boundary
of (\ref{cond2}), encountering a region in which neither Euler equation
has a solution and the density is fixed at $n_i=1$. Further out a solution
of (\ref{cond1}) becomes possible, and for $i>39$ even the low-density
Euler equation ceases to have a solution, implying $n_i=0$. 
Density profiles for $U=4$ and different fillings are shown in Fig.~\ref{fig2}.
Each of the data sets in Fig.~\ref{fig2} illustrates one of the five phases
of the $U>0$ part of the phase diagram, Fig.~\ref{fig3}. These are the same
five phases also obtained numerically in Refs.~\cite{drummond,rigol}. 

The present TLDA approach localizes the sites with $n_i=1$ by the conditions 
that $0<n<2$ and $\mu$ such that neither Euler equation has a solution.
On the other hand, in a Kohn-Sham approach \cite{polini,balda,gs,gsn} a
self-consistent solution is obtained iteratively, yielding in each iteration
step a density that is defined simultaneously at all sites. If for particular 
sites there is no value of $n_i$ corresponding to a solution of the Euler 
equations, the Kohn-Sham self-consistency cycle does not find any solution 
at all and does not converge. This is precisely what was observed in trying 
to perform calculations of the type of Refs.~\cite{balda,polini} in a region 
with plateaus. The TLDA procedure, which allows to obtain and characterize 
solutions site by site, has no such problem.

The negative $U$ analysis is performed in the same way, by 
starting from Eq.~(\ref{euler3}). Since there is only one branch, the
criterium for a valid solution is
\be
\frac{U}{2} + V_i - 2\pi I(|U|) \le \mu \le  \frac{U}{2} + V_i + 2\pi I(|U|).
\ee
For $\mu$ inside this interval the solution of (\ref{euler3}) is 
$n_i = (2/\pi)\arccos\left((V_i + U/2 - \mu)/(2\pi I(|U|))\right)$.
For $\mu < \frac{U}{2} + V_i - 2\pi I(|U|)$ the solution
is $n_i=0$ and for $\mu > \frac{U}{2} + V_i + 2\pi I(|U|)$ it is $n_i=2$. 
Typical density profiles in this region (not shown) are of the same type as
for $U>0$, with possible plateaus at $n=0$ and $n=2$. The absence of any
possible plateaus with $n=1$ is due to
the absence of a derivative discontinuity of $E_c$. The atomic density waves 
found from Kohn-Sham calculations in Ref.~\cite{polini} are not reproduced by 
the TLDA, showing that observation of these oscillations (just as Friedel 
oscillations) require an exact (orbital) treatment of the single-body kinetic 
energy.

The expresions for the phase boundaries obtained within TLDA can be 
used to analytically construct the full $U$-$\mu$ phase diagram, shown in 
Fig.~\ref{fig3}. For a harmonic confining potential of the form $V_i=k i^2$
region II (characterized by a central plateau with $n=1$, i.e., a local
Mott-insulator-like state) is bounded by 
$k - 2\cos(\pi/\beta) < \mu < U + 2\cos(\pi/\beta)$. 
Region III (two lateral plateaus with $n=1$) is bounded by 
$U+2\cos(\pi/\beta)<\mu<k + U/2 + (U/4 + \cos(\pi/\beta))^2/k$,
and region V (a central band-insulator-like plateau with $n=2$) by 
$\mu > U + 2 + k$. 
Region IV is the set of points satisfying simultaneously the criteria for 
region III and V, i.e., has lateral plateaus with $n=1$ and a central plateau 
with $n=2$.
Region I (purely metallic without any plateaus, except for, possibly, sites 
with $n=0$ at the wings of the trap) is the set of points with $\mu>-2$ 
that do not belong to any other $U>0$ region. 
In the TFA ($E_c\equiv 0$) the $U>0$ phases II, III and IV disappear, and the
dividing line between phase I and V becomes $\mu = U + 2 + k$. Phases II, III
and IV thus owe their existence to correlation.
Region VI is the $U<0$ extension of region V and characterized by a 
plateau at $n=2$ in the trap center. It is bounded from below by 
$\mu=k+U/2+2\pi I(|U|)$.
Region VII, finally, is the $U<0$ extension of region I and characterized
by the absence of any plateaus except for, possibly, sites with $n=0$ at the 
wings of the trap. 

Special values of $U$ and $\mu$ can be used to further classify the 
possibilities: For $\mu<\mu^*$, where $\mu^*(U>0)=-2$ and 
$\mu^*(U<0)=  U/2-2 \pi I(|U|)$, the only solution is $n_i=0$ at all sites, 
i.e., an empty system (hatched region in the phase diagram).
For $0<U<U^*$ the intervals permitting solutions of (\ref{cond1}) and 
(\ref{cond2}) overlap, and no $n=1$ plateaus can appear, leaving only
phases I, V and VI. (In this situation, at isolated sites, an $n>1$ 
and an $n<1$ solution can coexist.) For $U^*<U<U^{**}(k)$, where $U^{**}(k)$ 
is the solution of $(U/4 + \cos(\pi/\beta))^2=k(U/2+2)$, there cannot be a 
phase IV (i.e, plateaus at $n=1$ {\em and} $n=2$) and instead there appears a
reentrant metallic phase of type I above phase III.

In summary, we have constructed the complete $U$-$\mu$ phase diagram of
one-dimensional
interacting harmonically confined fermions. For $U>0$ we confirm earlier
numerical results, but complement them by providing analytical expressions
for the phase boundaries. The TLDA approach also provides an explanation
why Kohn-Sham calculations do not work in the metal Mott-insulator 
phase-separated state (with plateaus at $n=1$) and provides an alternative way 
for obtaining density profiles in this region. For $U<0$ we propose, in
Eq.~(\ref{uneg}), a simple analytical parametrization of the Bethe-Ansatz 
solution, which we use, within TLDA, to extend the phase diagram to attractive
interactions. 

{\bf Acknowledgments}\\
This work was sup\-por\-ted by FAPESP and CNPq. We thank Mariana Odashima 
for useful discussions on the Bethe-Ansatz parametrization for $U<0$ and
Marco Polini for useful discussions on the physics at $U<0$.

\end{document}